\documentclass[10pt,twocolumn,letterpaper]{article}

\usepackage{cvpr}
\usepackage{times}
\usepackage{epsfig}
\usepackage{graphicx}
\usepackage{amsmath}
\usepackage{amssymb}
\usepackage{booktabs}
\usepackage[breaklinks=true,bookmarks=false]{hyperref}

\usepackage[breaklinks=true,bookmarks=false]{hyperref}

\cvprfinalcopy 

\def\cvprPaperID{****} 

\begin{document}

\title{Real Time Animator: High-Quality Cartoon Style Transfer in 6 Animation Styles on Images and Videos}

\author{Liuxin Yang\\
Electrical Engineering\\
{\tt\small lyang822@stanford.edu}
\and
Priyanka Ladha \\
Computer Science(AI), Graduate School of Business  \\ 
{\tt\small priladha@stanford.edu}
}

\maketitle

\begin{abstract}
This paper presents a comprehensive pipeline that integrates state-of-the-art techniques to achieve high-quality cartoon style transfer for educational images and videos. The proposed approach combines the Inversion-based Style Transfer (InST) framework for both image and video style stylization, the Pre-Trained Image Processing Transformer (IPT) for post-denoising, and the Domain-Calibrated Translation Network (DCT-Net) for more consistent video style transfer. By fine-tuning InST with specific cartoon styles, applying IPT for artifact reduction, and leveraging DCT-Net for temporal consistency, the pipeline generates visually appealing and educationally effective stylized content. Extensive experiments and evaluations using the scenery and monuments dataset demonstrate the superiority of the proposed approach in terms of style transfer accuracy, content preservation, and visual quality compared to the baseline method, AdaAttN. The CLIP similarity scores further validate the effectiveness of InST in capturing style attributes while maintaining semantic content. The proposed pipeline streamlines the creation of engaging educational content, empowering educators and content creators to produce visually captivating and informative materials efficiently. 
\end{abstract}

\section{Introduction}

Cartoon style transfer has emerged as a powerful technique for creating visually appealing and engaging educational content. However, existing methods often face challenges in preserving semantic information, maintaining temporal consistency in videos, and dealing with artifacts introduced during the stylization process. Addressing these issues is crucial for generating high-quality educational materials that effectively convey knowledge while captivating learners.

In this work, we present a comprehensive pipeline that combines the strengths of multiple state-of-the-art techniques to achieve high-quality cartoon style transfer for educational images and videos. Our pipeline consists of three key components: fine-tuning the InST framework with particular styles for images and videos (via frames clipping), applying post-denoising using the IPT, and integrating the DCT-Net for video style transfer. Figure \ref{Overview} shows our project design.

First, we fine-tune the InST framework using the Japanese anime style dataset, enabling the model to learn and capture the unique artistic characteristics of the desired style. This fine-tuning process allows us to generate stylized images that accurately reflect the chosen cartoon aesthetics while preserving the semantic content of the educational materials.

Next, we introduce a post-denoising step using the pre-trained IPT model. The IPT model, known for its exceptional denoising capabilities, is employed to mitigate artifacts and noise that may arise during the style transfer process. By leveraging the transformer-based architecture of IPT, we can effectively remove distortions and enhance the visual quality of the stylized images, resulting in cleaner and more visually appealing educational content.

Finally, we explore the video style transfer using both InST model and DCT-Net. While we extend the InST model to videos, we also explore DCT-Net, a state-of-the-art framework specifically designed for video style transfer.  DCT-Net ensures temporal consistency and coherence across frames. 

The integration of these three components into a comprehensive pipeline sets our approach apart from existing methods. By leveraging the strengths of InST for image and video style transfer, IPT for post-denoising, and DCT-Net for video style transfer, we can generate high-quality educational materials that exhibit consistent artistic styles, improved visual clarity, and enhanced temporal coherence. We demonstrate the effectiveness of our pipeline through extensive experiments and evaluations using the scenery and monuments dataset. The generated stylized images and videos showcase the successful transfer of cartoon styles while preserving the essential educational content. Our results highlight the potential of our approach in creating visually captivating and educationally effective materials that engage and inspire learners.

\section{Related Work}

To enable style transfer on the image level, we employ the InST framework with diffusion models for artistic style transfer \cite{zhang2023inversion}. It learns the style representation of a reference image through inversion and guides the synthesis of stylized images. InST achieves impressive results in transferring various artistic styles while preserving the content of the source images. Our work builds upon the concepts introduced in InST, particularly the use of diffusion models for style transfer. However, instead of directly extending InST to the video domain, we incorporate a separate framework, DCT-Net \cite{men2022dct}, specifically designed for video style transfer.

DCT-Net is a novel image translation architecture proposed by Men et al \cite{men2022dct}. for few-shot portrait stylization. It consists of a content adapter, a geometry expansion module, and a texture translation module. DCT-Net addresses the challenges of limited style exemplars and achieves high-quality stylization results with advanced content preservation and handling of complicated scenes. In our work, we leverage DCT-Net to enable video style transfer for educational content. While DCT-Net is originally designed for portrait stylization, we adapt it to handle educational videos. By incorporating DCT-Net into our pipeline, we ensure temporal consistency and preserve the essential educational content throughout the stylized videos.

Our InST framework for image style transfer shows superior results in terms of the evaluation metric (CLIP similarity) \cite{peng2006clip} compared to the baseline model, which is the Adaptive Attention Normalization (AdaAttN) \cite{liu2021adaattn}.
AdaAttN is a module designed for style transfer that combines attention and normalization. It aims to achieve a better balance between content preservation and style transfer by considering features from shallow to deep layers of both the content and style images. While AdaAttN achieves excellent results in arbitrary style transfer, InST demonstrates superior performance by leveraging diffusion models, inversion-based style representation, and stochastic inversion. 

In our work, we draw inspiration from IPT and incorporate a denoising module into our video style transfer pipeline. IPT is a pre-trained model that leverages the transformer architecture for image processing tasks such as denoising, super-resolution, and deblurring \cite{chen2021pre}. It learns to capture intrinsic features and transformations from a large-scale dataset of corrupted image pairs. In our work, we adapt its concepts to develop a denoising component that enhances the visual quality of the stylized educational videos by reducing artifacts and noise introduced during the stylization process.

\section{Data}
For fine-tuning the Inversion-based Style Transfer (InST) framework, we utilize the (\hyperlink{https://www.kaggle.com/datasets/weiwangk/japanese-anime-scenes}{anime scenery dataset}), which consists of a diverse collection of Japanese anime scenes. This dataset contains a rich variety of artistic styles and scenery images, making it suitable for learning and capturing the unique characteristics of Japanese anime aesthetics.

The InST framework is trained on a large-scale dataset of image pairs, consisting of content images from the COCO dataset and style images from various artistic sources. This ensures the generalization capability of the InST model across a wide range of styles.

The Image Processing Transformer (IPT) model, which we employ for post-denoising, is pre-trained on a large-scale dataset derived from ImageNet, with algorithmically generated corrupted image pairs, as mentioned in the original paper. This allows IPT to learn intrinsic features and transformations for effective image restoration and denoising.

The anime scenery dataset used for fine-tuning InST contains a sufficient number of images to capture the diversity of Japanese anime styles. The InST and IPT models are trained on large-scale datasets, ensuring robustness and generalization capabilities \cite{zhang2023inversion}, \cite{chen2021pre}. By leveraging these datasets, our cartoon style transfer pipeline is well-equipped to handle a wide range of educational images and videos, delivering high-quality stylized results.

For the use of the DCT-Net, our testing videos come from this websit: \hyperlink{https://www.pexels.com/search/videos/monuments/}{Pexels}. In Pexels, we find an extensive collection of short video clips related to monuments, such as the Statue of Liberty and the Eiffel Tower,and people visiting and exploring these iconic structures. Additionally, the website features captivating clips of historic European castles, churches, and other architecturally significant buildings that serve as monuments. If we convert those original videos into catoonized videos, we can help children better learn about the world-famous monuments and scenary.

\section{Methods}

To address the challenges of cartoon style transfer for educational images, we propose a comprehensive pipeline that combines the strengths of the InST framework, the IPT model, For the style transfer for educational videos, we explore two methods. The first method is clipping videos into frames and feed the frames into the InST, where InST will process them separately. Then we will combine the generated images together to form the final styled videos. The second method is using the DCT-Net, specifically designed for video style transfer.

\subsection{Inversion-Based Style Transfer with Diffusion Models for Images}

The InST framework serves as the foundation for our style transfer approach. InST utilizes diffusion models to learn the style representation of a reference image through an inversion process. The framework consists of three key components: style inversion, conditional synthesis, and stochastic inversion.

The style inversion process in InST employs an attention-based inversion module to learn the textual embedding of the reference style image. This module takes the CLIP image embedding of the reference image as input and applies multi-layer cross-attention to extract the key style information. The cross-attention mechanism allows the module to attend to different regions of the image embedding and capture the essential characteristics of the artistic style. By learning a compact textual embedding, the style inversion process effectively encodes the unique style attributes of the reference image, enabling accurate style transfer.

The conditional synthesis process uses the textual embedding learned from the style inversion to guide the diffusion model during image synthesis. The textual embedding is concatenated with the noisy input at each timestep of the diffusion process, providing a consistent style conditioning throughout the image generation. The diffusion model, which is trained to denoise the noisy input and generate realistic images, is influenced by the style embedding to produce images that possess the desired artistic style. At the same time, the content of the source image is preserved by conditioning the diffusion model on the original image content. This ensures that the generated stylized images maintain the semantic information present in the source image while adopting the characteristics of the reference style.

To maintain semantic consistency between the content image and the stylized output, InST introduces a stochastic inversion process. This process involves adding random Gaussian noise to the content image at different scales and using the denoising U-Net of the diffusion model to predict the noise. The denoising U-Net is trained to estimate the noise present in the input image, effectively learning to separate the content from the style. By predicting the noise and subtracting it from the noisy input, the denoising U-Net recovers the content information while discarding the style-related noise. The predicted noise is then used as the initial input during the synthesis stage, providing a content-preserving starting point for the diffusion model. This stochastic inversion process helps to retain the semantic content of the source image in the stylized output, ensuring that the important educational information remains intact.

\subsection{Denoising}
In our InST framework, denoising can be performed at different stages to enhance the quality and smoothness of the generated stylized images. We explore several approaches to incorporate denoising into the InST pipeline:

\begin{itemize}
    \item Denoising as part of the generative model during training:
InST utilizes latent diffusion models (LDMs) as the backbone for both the inversion process and the generation of stylized images. LDMs learn to generate smooth and natural images by iteratively adding Gaussian noise to the image or its latent representation and then learning the reverse denoising process. 
    \item Denoising as an intermediate step during generation:
InST introduces a novel stochastic inversion module that incorporates denoising as an intermediate step during the image generation process. The stochastic inversion module first adds random noise to the content image and then uses the denoising U-Net of the diffusion model to predict and remove the noise. The predicted noise is then used as the initial input noise for the subsequent generation steps. 
    \item Denoising as a post-processing step:
After the stylized image is generated, traditional denoising algorithms or specially trained denoising networks can be applied as a post-processing step. These methods aim to reduce noise, JPEG compression artifacts, and other distortions in the generated image. 
\end{itemize}
\subsubsection{Pre-Trained Image Processing Transformer for Post
Image Denoising}

To address the issue of artifacts and noise introduced during the style transfer process, we incorporate the IPT as a post-denoising step in our pipeline. The IPT model, proposed by Chen et al. \cite{chen2021pre}, leverages the power of transformer architectures to effectively remove noise and artifacts from images.

The IPT model is pretrained on a large-scale dataset of corrupted image pairs, where each pair consists of a clean image and its corresponding corrupted version. The dataset is derived from the ImageNet dataset and includes various types of image degradation, such as Gaussian noise, motion blur, and JPEG compression. The pre-training process allows the IPT model to learn intrinsic features and transformations that are effective for image restoration and denoising.

The architecture of the IPT model follows an encoder-decoder structure, with the transformer blocks serving as the backbone. The encoder takes the corrupted image as input and applies multi-head self-attention mechanisms to capture global dependencies and extract meaningful features. The self-attention mechanism allows the model to attend to different regions of the image and capture long-range dependencies, which is crucial for effective denoising. The decoder then uses cross-attention to attend to the encoded features and generates the denoised output image. The transformer architecture enables the IPT model to learn complex image restoration transformations and effectively remove noise and artifacts.

\subsection{Baseline: AdaAttN: Revisit Attention Mechanism in Arbitrary Neural Style Transfer}

AdaAttN is designed for style transfer that combines attention and normalization. AdaAttN aims to achieve a better balance between content preservation and style transfer by considering features from shallow to deep layers of both the content and style images. It computes attention maps using these features and utilizes the obtained attention maps to calculate weighted mean and standard variance feature maps of the style features. AdaAttN then performs adaptive normalization, where the content features are normalized to align their local feature statistics with the computed style feature statistics on a per-point basis.

While AdaAttN achieves excellent results in arbitrary style transfer, the InST method demonstrates superior performance for several reasons. First, it utilizes diffusion models, which have shown remarkable performance in image synthesis tasks, enabling the generation of high-quality stylized images with diverse and visually appealing results. Second, the inversion-based approach of InST allows it to capture the essential characteristics of the artistic style from a single reference image, making it more practical and efficient for transferring styles from limited reference images. Finally, InST introduces a stochastic inversion process that helps maintain semantic consistency between the content image and the stylized output, preserving the important semantic information present in the content image.

\subsection{Video Style Transfer}

\subsubsection{Inversion-Based Style Transfer with Diffusion
Models for Videos}

To extend the InST framework for video style transfer, we first extract individual frames from the input video. Each frame is then processed independently using the InST model, applying the same style transfer technique as used for image style transfer. This involves style inversion, conditional synthesis, and stochastic inversion to generate stylized frames that maintain the semantic content of the original frames while adopting the desired artistic style. Finally, the generated stylized frames are combined to reconstruct the complete video, resulting in a stylized educational video that preserves the original content with the applied artistic style.

\subsubsection{DCT-Net: Domain-Calibrated Translation for Portrait Stylization}

To alleviate the consistency problems in using InST for video style transfer, we settled down to the DCT-Net. DCT-Net is a novel image translation architecture designed for few-shot portrait stylization that ensures temporal consistency and coherence across video frames.

Using DCT-Net can achieve temporally consistent and visually appealing stylized videos. The content adapter helps in calibrating the content distribution of the target domain, ensuring that the stylized frames maintain coherence with the original video content. The geometry expansion module enhances the spatial flexibility of the stylization process, allowing for adaptive deformations and preserving the facial structure across frames. The texture translation module learns a fine-grained texture mapping, resulting in stylized videos with detailed and consistent artistic patterns.

During inference, the video frames are processed sequentially by the trained model, and the stylized frames are combined to generate the final stylized video. The incorporation of DCT-Net ensures that the stylized video maintains temporal coherence and exhibits smooth transitions between frames, avoiding flickering or sudden changes in style.

One important thing to mention is that we combine the weights from StyleGAN2-Official, StyleGAN2-Pytorch and DCT-Net. We include the details in the experiment section.

\section{Experiments}
\subsection{Post Image Denoising}
In our exploration of post-denoising techniques for enhancing the visual quality and clarity of the stylized outputs generated by InST, we investigated three different approaches:

\begin{itemize}
    \item OpenCV's Built-in Denoising Function:
We first experimented with OpenCV's built-in denoising function, specifically the \texttt{cv2.fastNlMeansDenoisingColored} function. This function applies the Fast Non-Local Means Denoising algorithm to the input image, effectively reducing noise while preserving edges and details. We applied this function to the stylized images generated by InST, specifying appropriate parameters such as the filter strength and color component weights. While this approach provided some level of noise reduction, we observed that it may not be optimal for handling the specific types of artifacts and distortions present in the stylized images. Given the generated anime image in Figure \ref{fig:my_image1}, the denoised image using OpenCV is shown in Figure \ref{07_denoised_opencv}. We can see that OpenCV's built in denoising function works well by alleviating some artifacts in the generated image.
\begin{figure}[h]
    \centering
    \includegraphics[width=0.35\textwidth]{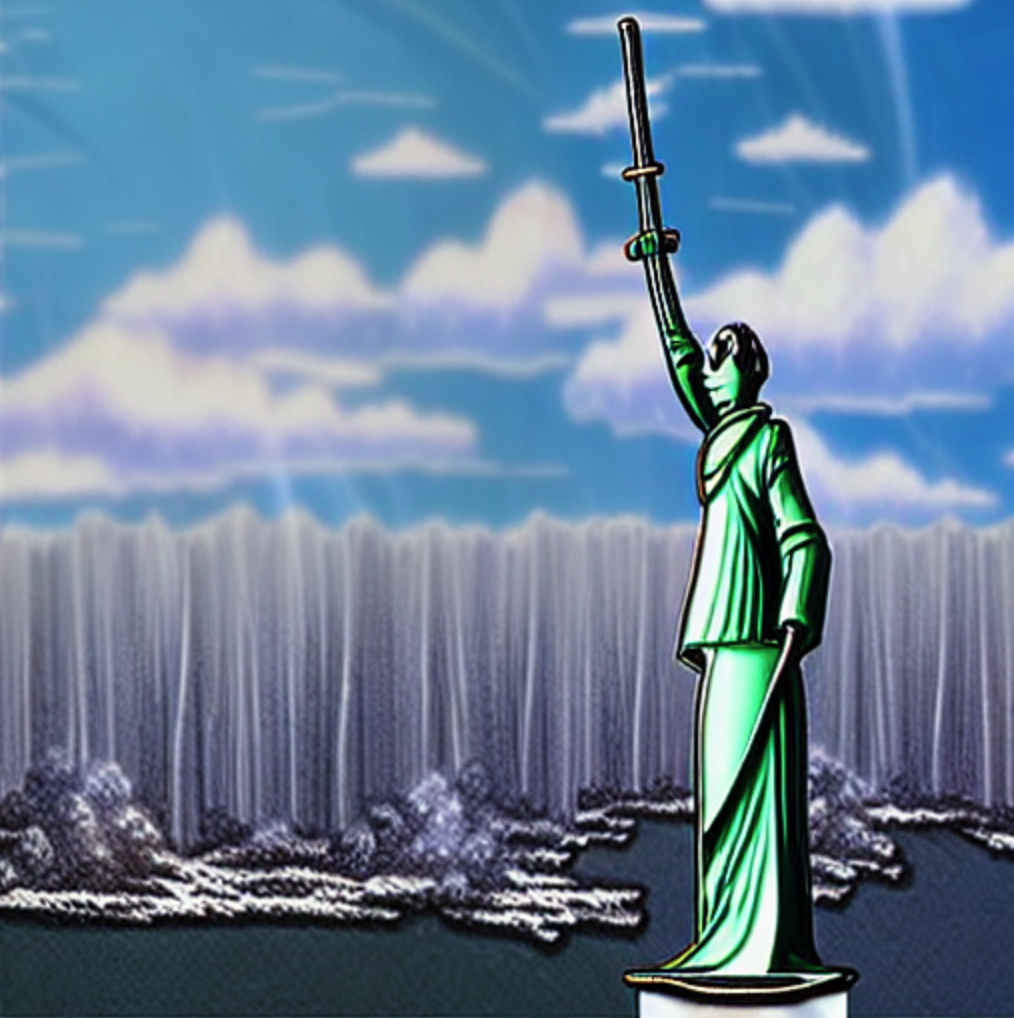}
    \caption{Generated Anime Image}
    \label{fig:my_image1}
\end{figure}
\begin{figure}[h]
    \centering
    \includegraphics[width=0.35\textwidth]{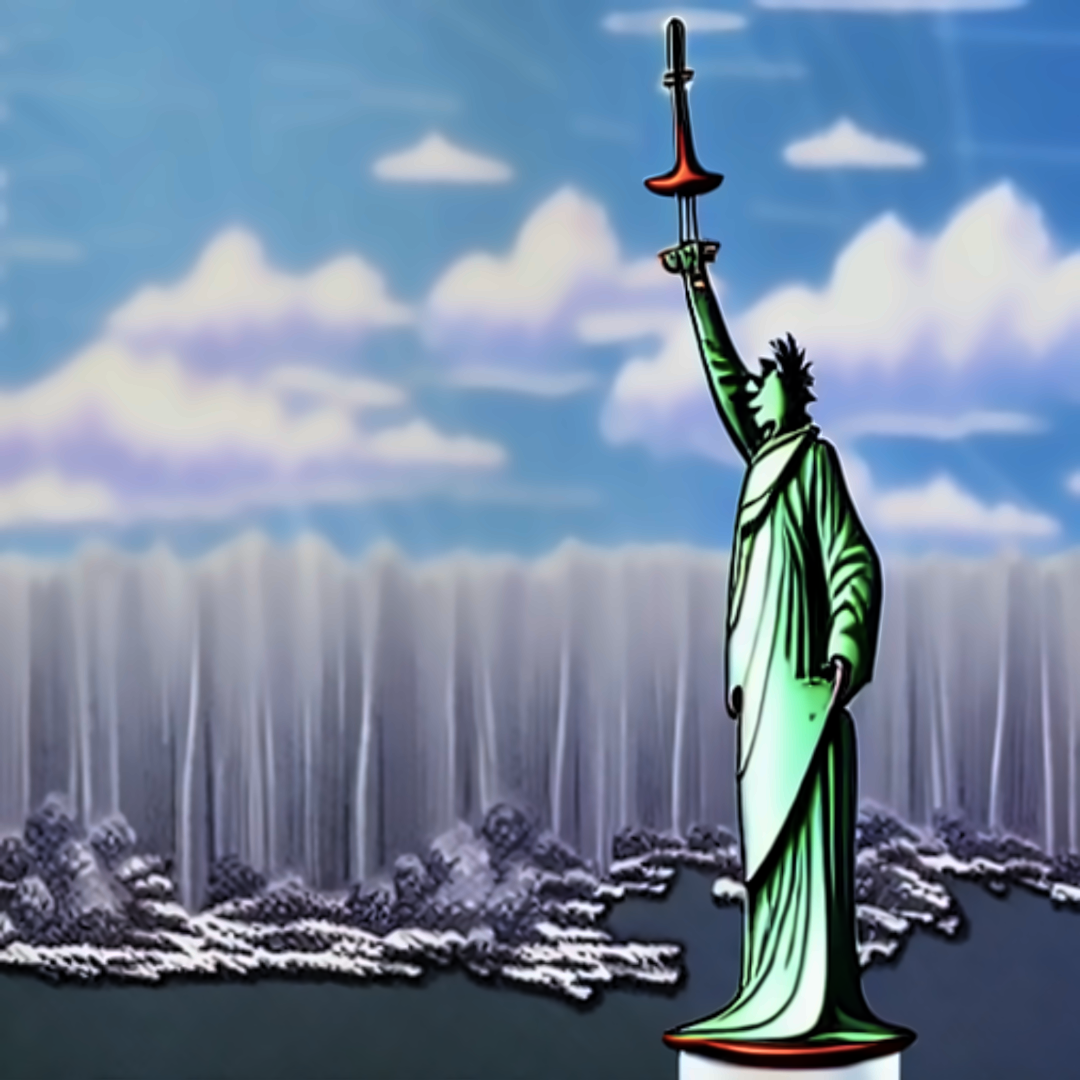}
    \caption{Denoised by OpenCV}
    \label{07_denoised_opencv}
\end{figure}
\item DnCNN (Denoising Convolutional Neural Network) \cite{zhang2017beyond}:
We also explored the use of DnCNN, a deep learning-based denoising model that has shown promising results in removing Gaussian noise from images. However, we encountered a limitation with the pre-trained DnCNN models available online. These models were trained specifically for denoising grayscale images, whereas the stylized images generated by InST are in color. Due to this incompatibility, we were unable to directly apply the pre-trained DnCNN models to our stylized images. Training a custom DnCNN model for color image denoising would require a significant amount of time and resources, which was beyond the scope of our current exploration.
\item IPT (Image Processing Transformer):
Finally, we investigated the use of IPT, a pretrained model that leverages the transformer architecture for various image processing tasks, including denoising. IPT has shown impressive results in removing noise and artifacts from images while preserving fine details and structures. We found that IPT was well-suited for our post-denoising needs, as it can handle color images and has been trained on a large-scale dataset of corrupted image pairs. By integrating IPT into our post-denoising pipeline, we were able to effectively reduce noise, JPEG compression artifacts, and other unwanted distortions in the stylized images generated by InST.

\end{itemize}

After evaluating the three approaches, we ultimately chose to incorporate IPT as our post-denoising solution due to its ability to handle color images, robustness in removing various types of artifacts, and pre-trained nature. However, we also face some challenges when integrating IPT to post denoise the generated images from the InST model. The pre-trained IPT model is designed to work with images of size 48×48 pixels, which is substantially smaller than the images generated by our InST model. This size limitation posed a problem, as directly applying IPT to our generated images would result in a significant loss of resolution and detail.

To address this issue, we explored various approaches to adapt IPT to our specific requirements. One potential solution was to resize the generated images to fit the input size of the IPT model. However, resizing the images to 48×48 pixels would result in an unacceptable level of quality degradation, as much of the fine detail and structure would be lost in the process.

After careful consideration, we devised a strategy to overcome this challenge. Instead of resizing the generated images to 48×48 pixels, we decided to resize them to a larger size of 96×96 pixels. This resizing step allowed us to preserve more of the original image detail while still being compatible with the IPT model.

To process the resized 96×96 images with IPT, we implemented a cropping and concatenation approach. We divided the resized image into four equal-sized patches of 48×48 pixels each. Each patch was then individually processed by the IPT model for denoising. After denoising, the four patches were concatenated back together to form the final denoised image.

While this approach allowed us to utilize IPT for post-denoising, it introduced a new challenge. The cropping and concatenation process resulted in visible seams or boundaries between the individual patches in the final denoised image. These artifacts were a consequence of the independent processing of each patch and the lack of seamless blending between them.

To mitigate the visibility of these seams, we experimented with various post-processing techniques. One approach involved applying a smoothing filter or a blending algorithm to the boundaries between the patches to create a more seamless transition. Another technique was to use overlapping patches during the cropping process and then blend the overlapping regions to minimize the appearance of seams.

Figure \ref{Resized} shows the resized 96*96 generated image from InST model. Figure \ref{merged_image} shows the denoised image using IPT. We can see that IPT works well in terms of denoising. Despite our efforts to minimize the visibility of the patch boundaries, the denoised images still exhibited some artifacts resulting from the cropping and concatenation process. This limitation highlights the trade-off between using a pre-trained model with a fixed input size and preserving the quality and consistency of the final denoised image.

\begin{figure}[h]
    \centering
    \begin{minipage}{0.22\textwidth}
        \centering
        \includegraphics[width=1\textwidth]{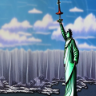}
        \caption{Resized 96*96 Image}
        \label{Resized}
    \end{minipage}
    \hfill
    \begin{minipage}{0.22\textwidth}
        \centering
        \includegraphics[width=\textwidth]{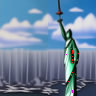}  
        \caption{Denoised Image Using IPT}
        \label{merged_image}
    \end{minipage}
    \caption{Post Image Denoising Using IPT}
    \label{fig:both_images}
\end{figure}

\subsection{Image Style Transfer Using Baseline}

\begin{figure}[h]
    \centering
    \includegraphics[width=0.35\textwidth]{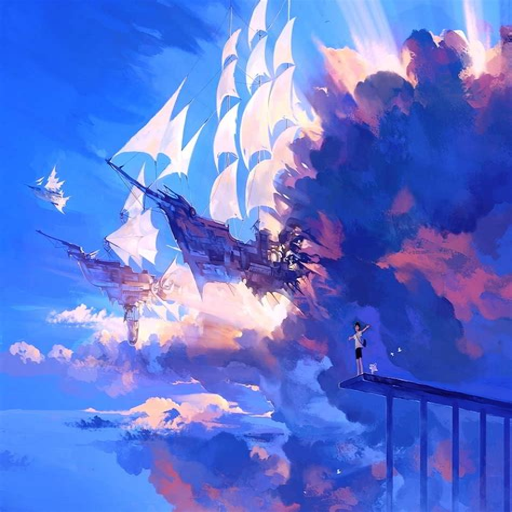}
    \caption{Style Image}
    \label{Style1}
\end{figure}

\begin{figure}[h]
    \centering
    \begin{minipage}{0.22\textwidth}
        \centering
        \includegraphics[width=1\textwidth]{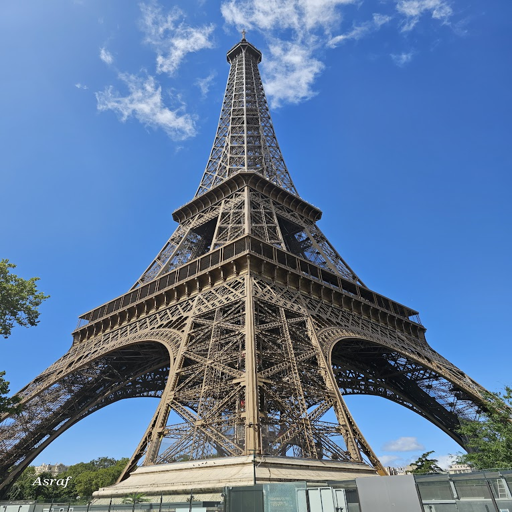}
        \caption{Content Image}
        \label{fig:baseline_c1}
    \end{minipage}
    \hfill
    \begin{minipage}{0.22\textwidth}
        \centering
        \includegraphics[width=\textwidth]{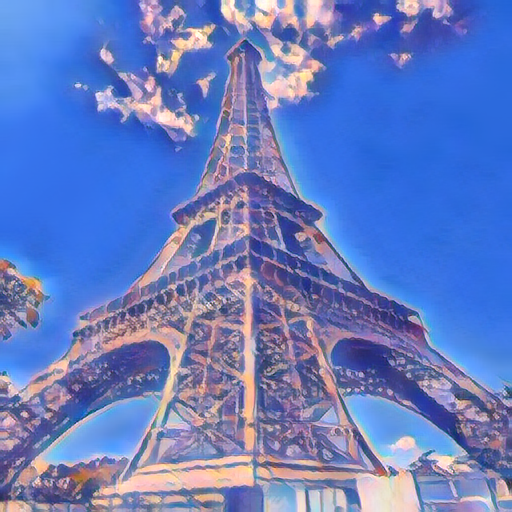}  
        \caption{Generated Image}
        \label{fig:baseline_g1}
    \end{minipage}
    \caption{Image Style Transfer Inference Using AdaAttN}
    \label{11}
\end{figure}

\begin{figure}[h]
    \centering
    \begin{minipage}{0.22\textwidth}
        \centering
        \includegraphics[width=1\textwidth]{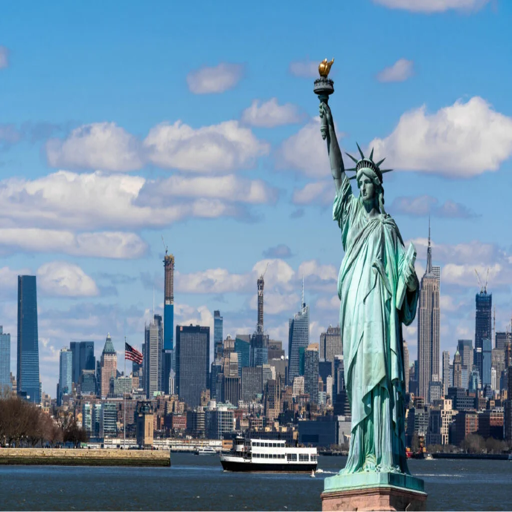}
        \caption{Content Image}
        \label{c2}
    \end{minipage}
    \hfill
    \begin{minipage}{0.22\textwidth}
        \centering
        \includegraphics[width=\textwidth]{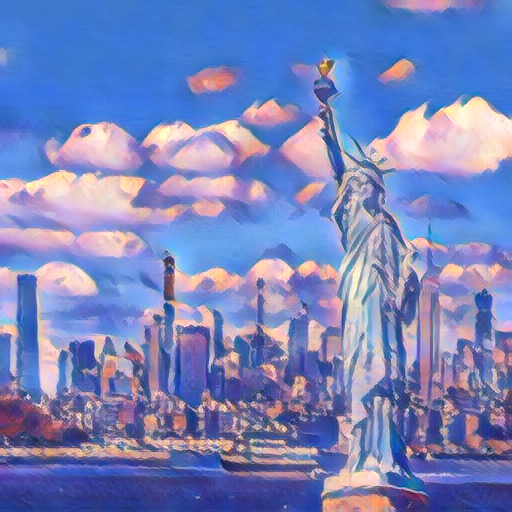}  
        \caption{Generated Image}
        \label{g2}
    \end{minipage}
    \caption{Image Style Transfer Inference Using AdaAttN}
    \label{12}
\end{figure}

\begin{figure}[h]
    \centering
    \begin{minipage}{0.22\textwidth}
        \centering
        \includegraphics[width=1\textwidth]{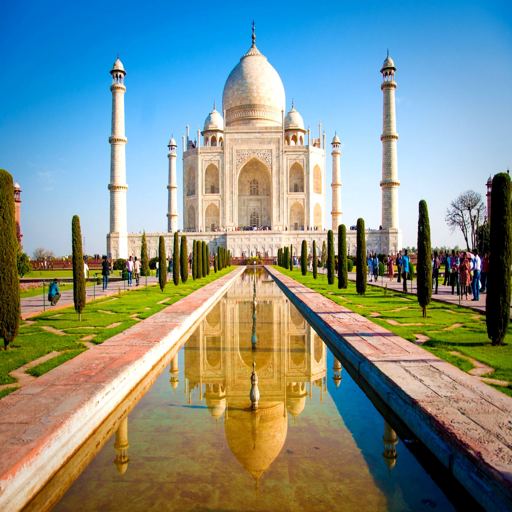}
        \caption{Content Image}
        \label{c3}
    \end{minipage}
    \hfill
    \begin{minipage}{0.22\textwidth}
        \centering
        \includegraphics[width=\textwidth]{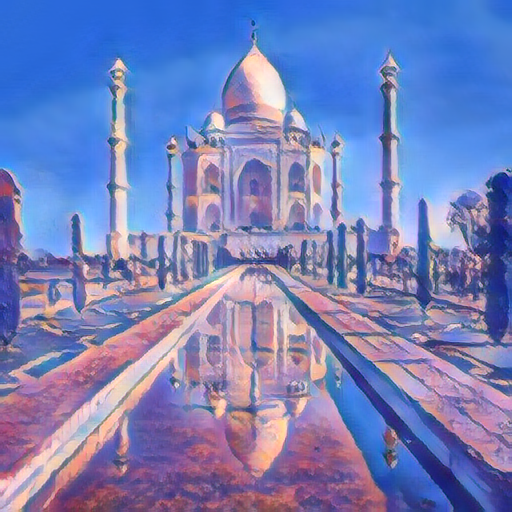}  
        \caption{Generated Image}
        \label{g3}
    \end{minipage}
    \caption{Image Style Transfer Inference Using AdaAttN}
    \label{13}
\end{figure}

Given the style image in Figure \ref{Style1} and different content images, we have the following generated images (see Figure \ref{11}, \ref{12}, and \ref{13}) using the baseline - AdaAttN.

\subsection{Image Style Transfer Using InST}

\begin{figure}[h]
    \centering
    \begin{minipage}{0.22\textwidth}
        \centering
        \includegraphics[width=1\textwidth]{th-384292406_1_c.png}
        \caption{Content Image}
        \label{fig:inst_c1}
    \end{minipage}
    \hfill
    \begin{minipage}{0.22\textwidth}
        \centering
        \includegraphics[width=\textwidth]{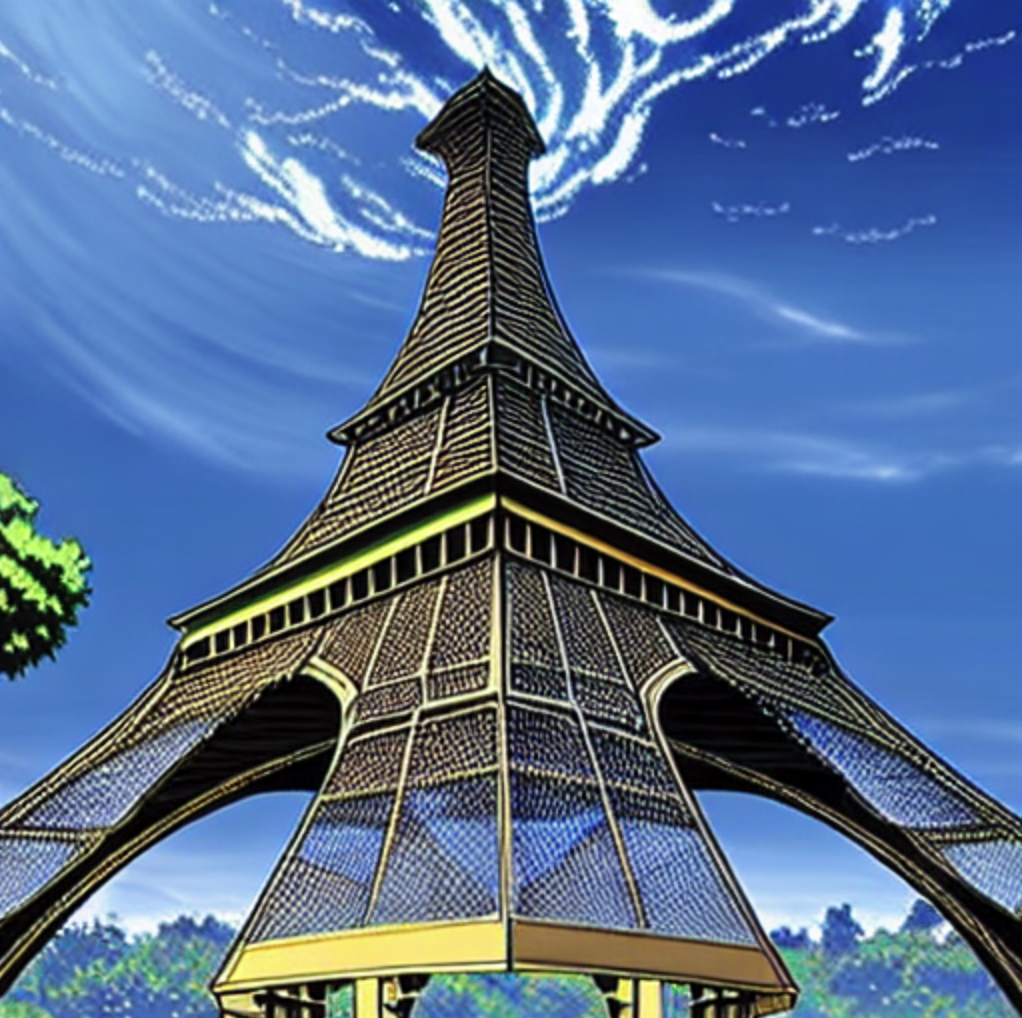}  
        \caption{Generated Image}
        \label{fig:inst_g1}
    \end{minipage}
    \caption{Image Style Transfer Inference Using InST}
    \label{21}
\end{figure}

\begin{figure}[h]
    \centering
    \begin{minipage}{0.22\textwidth}
        \centering
        \includegraphics[width=1\textwidth]{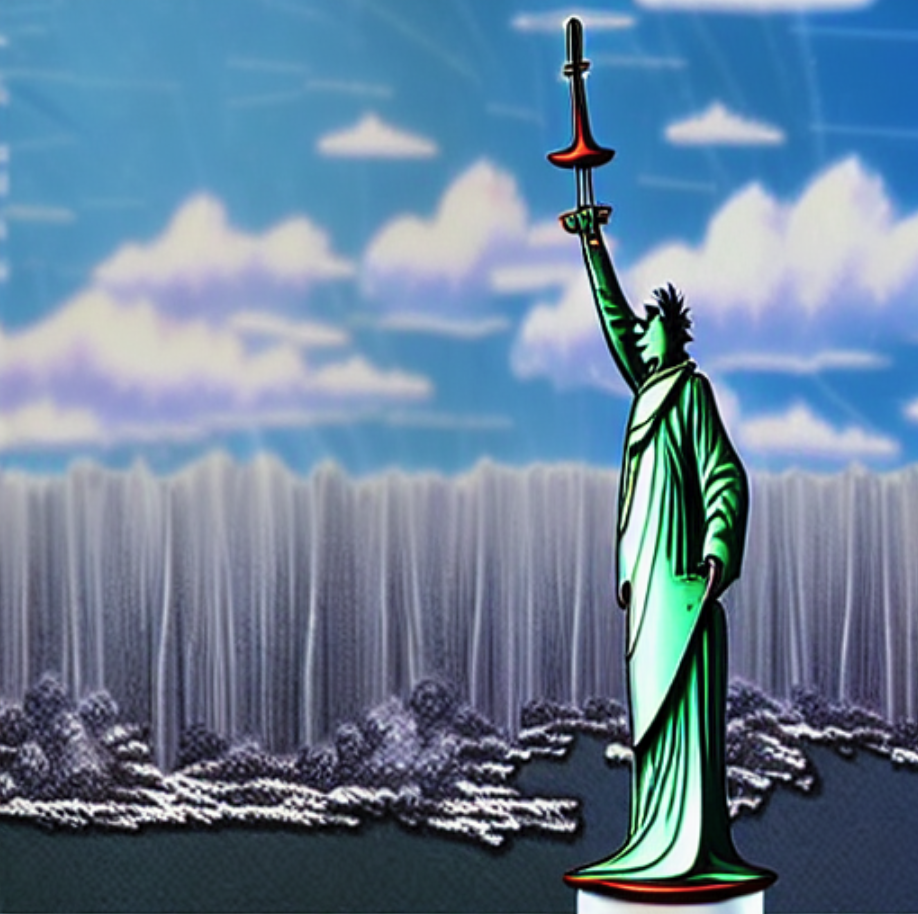}
        \caption{Content Image}
        \label{c2}
    \end{minipage}
    \hfill
    \begin{minipage}{0.22\textwidth}
        \centering
        \includegraphics[width=\textwidth]{Snipaste_2024-06-05_14-44-10.png}  
        \caption{Generated Image}
        \label{g2}
    \end{minipage}
    \caption{Image Style Transfer Inference Using InST}
    \label{22}
\end{figure}

\begin{figure}[h]
    \centering
    \begin{minipage}{0.22\textwidth}
        \centering
        \includegraphics[width=1\textwidth]{th-384292406_3_c.png}
        \caption{Content Image}
        \label{c3}
    \end{minipage}
    \hfill
    \begin{minipage}{0.22\textwidth}
        \centering
        \includegraphics[width=\textwidth]{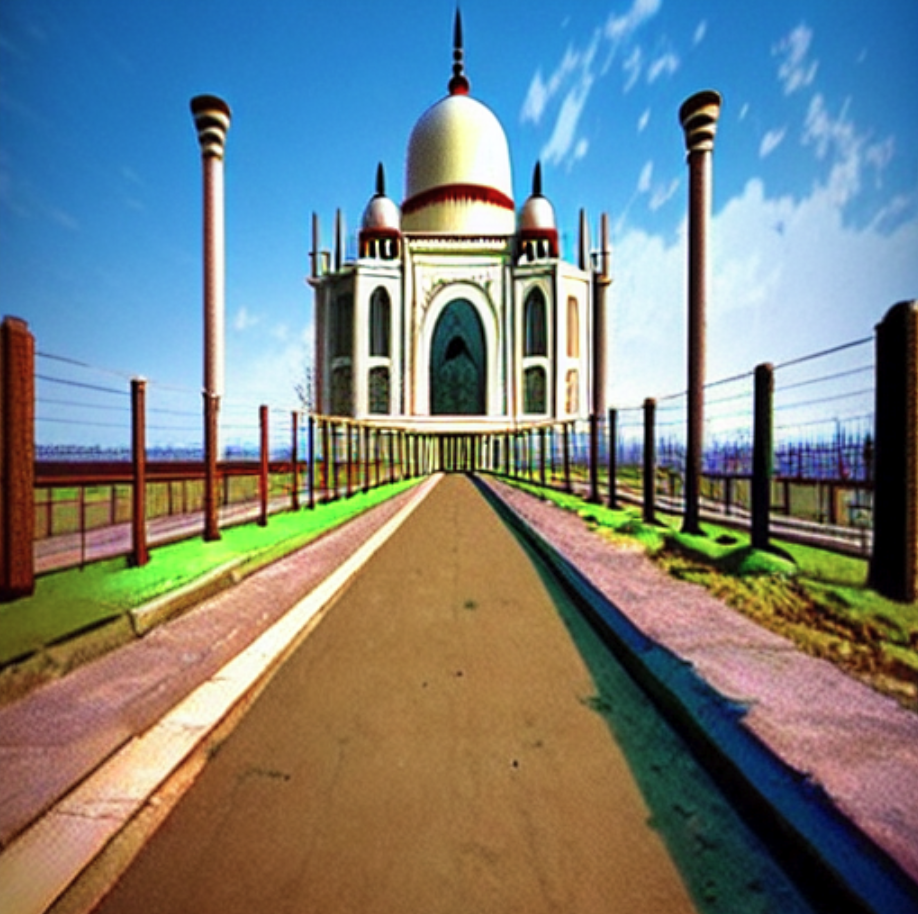}  
        \caption{Generated Image}
        \label{g3}
    \end{minipage}
    \caption{Image Style Transfer Inference Using InST}
    \label{23}
\end{figure}

Given the style image in Figure \ref{Style1} and different content images, we have the following generated images (see Figure \ref{21}, \ref{22}, and \ref{23}) using the InST model.

\subsection{Video Style Transfer Using InST}

When we apply the InST model to video style transfer by processing each frame separately and then combining the stylized frames back into a video, the resulting video appears to be choppy and lacks smoothness (seen in the appendix video link). The motion in the video is not fluid, and there are noticeable discontinuities or abrupt changes between consecutive frames. This issue can be attributed to several factors, such as the lack of temporal consistency, absence of motion information, variation in style transfer across frames, and frame-by-frame processing approach.

The lack of temporal consistency is a significant contributor to the choppiness of the stylized video. Since each frame is processed independently by the InST model, there is no explicit consideration for the continuity and motion between adjacent frames. The style transfer is applied to each frame in isolation, without taking into account the temporal context of the video. Moreover, the absence of motion information in the InST model further exacerbates the issue. The model operates on individual frames and does not incorporate any motion estimation or optical flow techniques, resulting in artifacts or inconsistencies in regions with significant motion.

To address these challenges and improve the smoothness of the stylized video, we further explored the DCT-Net, which will be explained in the next section in detail.

\subsection{Video Style Transfer Using Multiple Pretrained Models Integration}

Initially, we experimented with MTCNN network for increasing the accuracy of face and landmark detection, then resorted to StyleGAN2, and default weights of DCT-Net and reverted back the changes. The reason for change is that MTCNN was incompatible with DCT-Net and needed a lot of code changes, with multiple errors \cite{xiang2017joint}. 
We combined the checkpoints of StyleGAN2 with DCT-Net. We integrated \hyperlink{https://github.com/rosinality/stylegan2-pytorch}{StyleGAN2-PyTorch} and \hyperlink{https://github.com/NVlabs/stylegan2}{StyleGAN2 Official} to enhance the performance of our style transfer model \cite{karras2020analyzing}. StyleGAN2-Pytorch has been trained extensively on monuments whereas StyleGAN2 official has an extensive dataset of faces. We combined both the models to do fine tuning on monuments and faces. This is because some of our videos that have monuments also have human faces. This integration allowed us to improve the image quality and style diversity in DCT-Net, resulting in more realistic and varied style transfers. The combined approach helped achieve more consistent and stable outputs, reducing artifacts and enhancing the overall robustness of the model.

\begin{figure}[h]
    \centering
    \begin{minipage}{0.22\textwidth}
        \centering
        \includegraphics[width=1\textwidth]{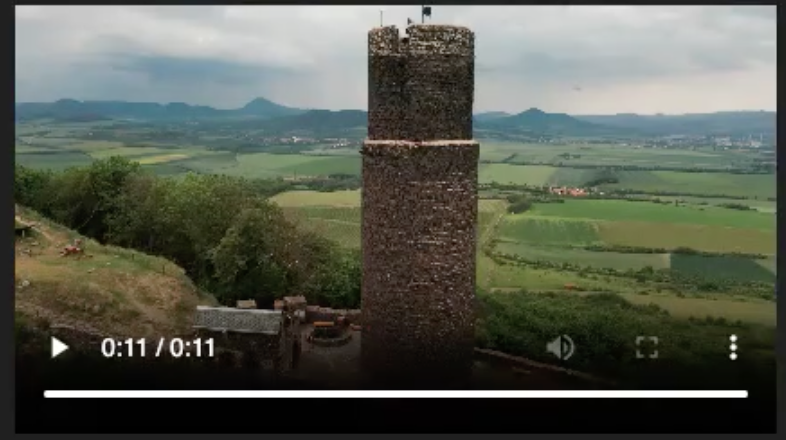}
        \caption{Original Video}
        \label{Resized}
    \end{minipage}
    \hfill
    \begin{minipage}{0.22\textwidth}
        \centering
        \includegraphics[width=\textwidth]{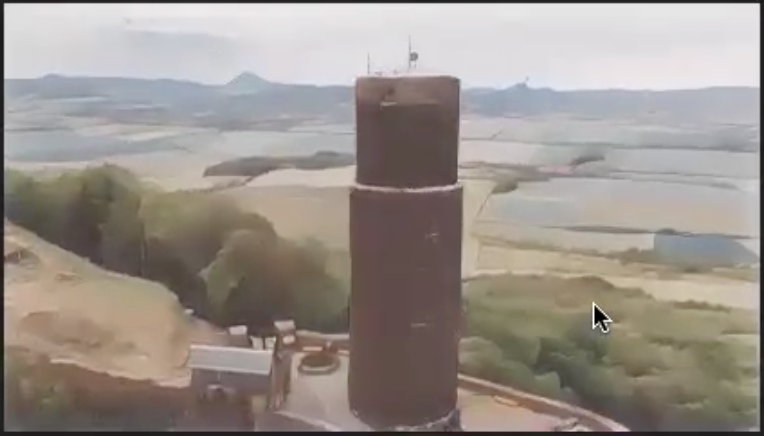}  
        \caption{Animated Video}
        \label{merged_image}
    \end{minipage}
      \centering
    \begin{minipage}{0.22\textwidth}
        \centering
        \includegraphics[width=1\textwidth]{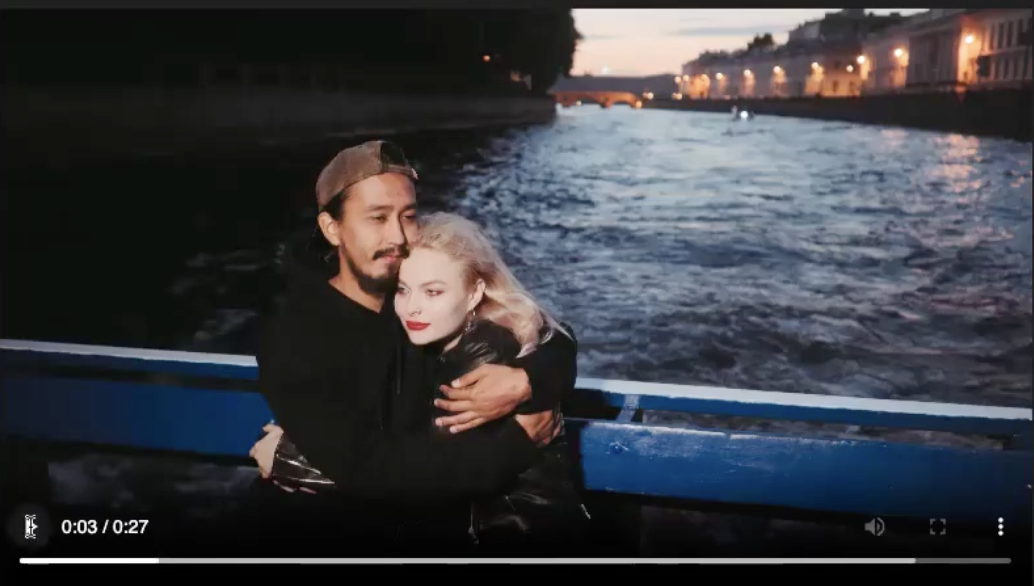}
        \caption{Original Video}
        \label{Resized}
    \end{minipage}
    \hfill
    \begin{minipage}{0.22\textwidth}
        \centering
        \includegraphics[width=\textwidth]{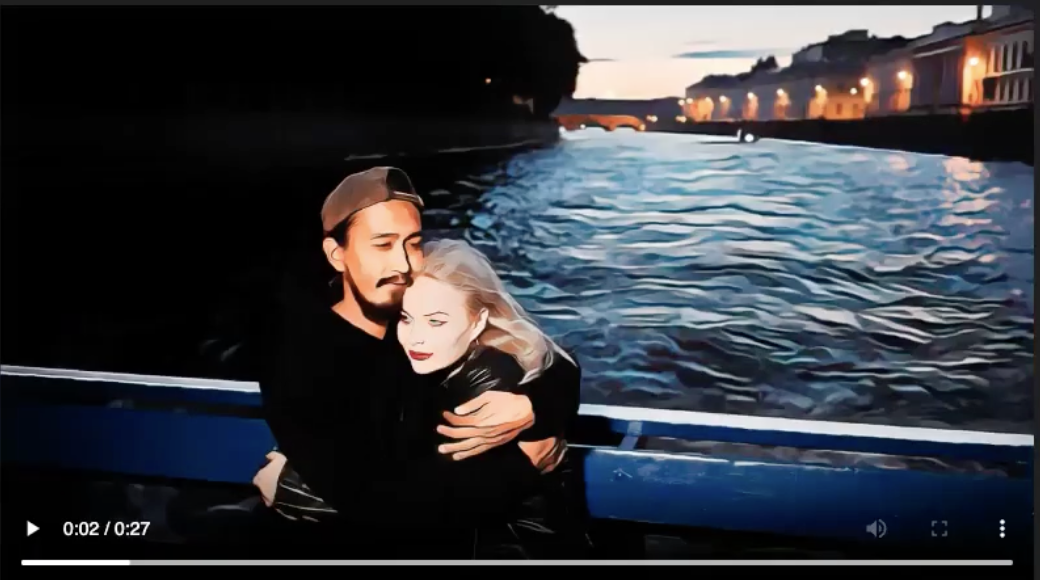}  
        \caption{Animated Video}
        \label{merged_image}
    \end{minipage}
    \caption{\href{https://drive.google.com/drive/folders/1m6fGOiY5ekwep3q7SwNuXaJAnVG_g5Lv?usp=sharing}{Video Results Link}}
    \label{fig:both_images}
\end{figure}

Our experiments show that if we only use the weights from the pretrained model, the generated video quality is much worse than combining both the weights from StyleGAN2 and DCT-Net. The detailed comparison between single weight source and combined weights is shown in our demo video.

The DCT-Net architecture and training process make it a promising approach for achieving better video style transfer results compared to other methods. The content calibration network (CCN) plays a crucial role in preserving content details and identity in the stylized video frames by using a pre-trained source generator to calibrate the content distribution of the target domain, ensuring content-symmetric features between the source and target domains. Additionally, the geometry expansion module (GEM) enhances the network's ability to handle variations in scale and rotation by applying affine transformations to both source and target samples, releasing spatially semantic constraints and enhancing geometry symmetry. This is particularly important for processing video frames with different orientations and sizes. Furthermore, the texture translation network (TTN) employs a U-Net architecture to learn fine-grained texture translation in the pixel level, enabling the network to capture local details and produce high-quality stylized video frames. The combination of these modules allows DCT-Net to generate stylized video frames that maintain content fidelity, handle variations in scale and orientation, and produce fine-grained texture details, making it a suitable approach for achieving high-quality video style transfer while preserving the content and temporal consistency of the original video.

\subsection{Evaluation}
To quantitatively assess the performance of InST in terms of style transfer and content preservation, we employ the CLIP (Contrastive Language-Image Pre-training) model to measure the similarity between the generated images and the corresponding style and content images. CLIP is a powerful visual-linguistic model that can calculate the similarity between a given image and a text description. By leveraging this capability, we utilize CLIP to evaluate how well the generated images retain the style attributes of the reference style image and the content of the original content image.

The CLIP similarity scores range from 0 to 1, with higher values indicating greater similarity. The high scores indicate that the generated images successfully capture the desired style attributes from the reference style image while maintaining the content of the original image. 

We use CLIP similarity scores as a quantitative metric to assess the performance of the proposed method in terms of style transfer and content preservation. From the results in Table 1 and Table 2, the high CLIP similarity scores between the generated images and both the style and content images demonstrate the effectiveness of InST in generating visually appealing and semantically meaningful stylized images.

\subsubsection{InST}

Table \ref{InST1} shows the CLIP similarity between the generated and style image, and also the CLIP similarity between the generated and content image using InST.

\begin{table}[h]
    \centering
    \setlength{\tabcolsep}{5pt}  
    \begin{tabular}{ccc}
        \toprule
        & \parbox{2.5cm}{\centering \textbf{Generated \& Style Img}} & \parbox{2.5cm}{\centering \textbf{Generated \& Content Img}} \\
        \midrule
        Img1 & 0.7012 & 0.8188 \\
        Img2 & 0.6308 & 0.7886 \\
        Img3 & 0.6904 & 0.6367 \\
        \bottomrule
    \end{tabular}
    \caption{InST}
    \label{InST1}
\end{table}

\subsubsection{Comparison with Baseline - AdaAttN}

Table \ref{AdaAttN1} shows the CLIP similarity between the generated and style image, and also the CLIP similarity between the generated and content image using AdaAttN.

\begin{table}[h]
    \centering
    \setlength{\tabcolsep}{3pt}  
    \begin{tabular}{ccc}
        \toprule
        & \parbox{3cm}{\centering \textbf{Generated \& Style Img}} & \parbox{3cm}{\centering \textbf{Generated \& Content Img}} \\
        \midrule
        Img1 & 0.6621 & 0.8140 \\
        Img2 & 0.6245 & 0.8053 \\
        Img3 & 0.6245 & 0.6348 \\
        \bottomrule
    \end{tabular}
    \caption{AdaAttN}
    \label{AdaAttN1}
\end{table}

\section{Conclusion}
In our project, we proposed a comprehensive pipeline that integrates InST, IPT, and DCT-Net to achieve high-quality cartoon style transfer for educational images and videos. The key results demonstrate the effectiveness of our approach in generating visually appealing and educationally effective stylized content. By fine-tuning InST with specific cartoon styles, we successfully captured the artistic characteristics while preserving the semantic content. We compare the performance of the InST and AdaAttN on the image style transfer. The incorporation of IPT as a post-denoising step significantly enhanced the visual quality by reducing artifacts and noise. Furthermore, we explore two methods for the video style transfer. One method is extend InST model to deal with videos and another method is to use the DCT-Net. DCT-Net works better than InST since it enables temporally consistent video style transfer, ensuring coherence and smooth transitions between frames.

The evaluation using CLIP similarity scores validated the superiority of InST over the baseline method, AdaAttN, in terms of style transfer accuracy and content preservation. The proposed pipeline achieved higher similarity scores between the generated images and both the style and content images, indicating its effectiveness in capturing style attributes while maintaining semantic information.

Through this work, we have learned the importance of leveraging state-of-the-art techniques and adaptively combining them to address the challenges of cartoon style transfer for educational content. The synergy between InST, IPT, and DCT-Net has proven to be a powerful approach for generating visually captivating and informative materials efficiently.

For future work, it would be interesting to explore the potential of replacing the widely used U-Net architecture in current diffusion-based image generation models with Vision Transformers (ViT). ViT has demonstrated superior performance in various computer vision tasks, especially in capturing long-range dependencies through self-attention mechanisms. By integrating ViT into the diffusion model framework, we aim to generate more globally coherent and detail-rich artistic images, enhance the model's adaptability to high-resolution inputs, and further improve the diversity and artisticness of the generated results. Key challenges to address include effectively incorporating ViT into the iterative refinement process of diffusion models, balancing performance and computational efficiency, and designing ViT-compatible artisticness loss functions. 

In conclusion, our work on cartoon style transfer for educational images using AdaAttN, InST and IPT. We also use InST and DCT-Net for videos. Our experiments demonstrated promising results and opened up new possibilities for creating engaging and effective educational content. 

\appendix

\section{First Appendix}
Video Results: \href{https://drive.google.com/drive/folders/1WYVy82ahT3uLk-OVHbsbkmwBgk_bQrGb?usp=sharing}{Link}
\begin{enumerate}
    \item \textbf{Video 1} : This video was generated using InST framework. As you can see video lacks temporal consistency. And thus we implemented DCT-Net for animating videos
    \item \textbf{Video 2} : This video was generated using DCT-Net. It compares the difference in the before and after quality of the videos wherein we used default weights of DCT-Net in before part and then updated the checkpoint of StyleGan2 to achieve much better quality.
    \item \textbf{Video 3} : This video, generated using DCT-Net shows the reason for combining StyleGan2-Pytorch and StyleGan2-Official. At times videos have both monuments and faces and thus we used weights from both these models to increase the video accuracy
\end{enumerate}

\section{Second Appendix}

\begin{figure}[h]
    \centering
\includegraphics[width=0.45\textwidth]{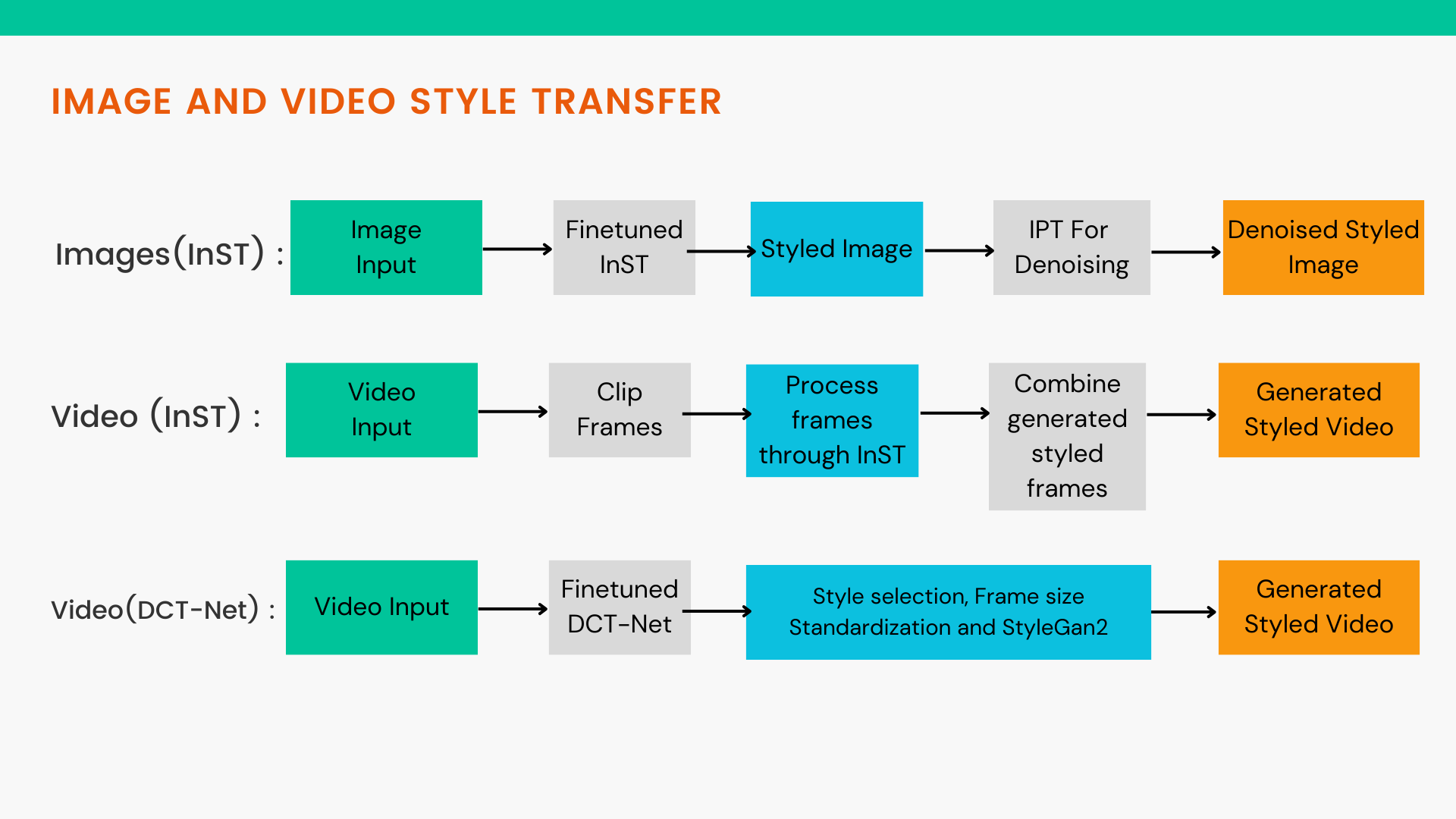}
    \caption{Overview}
    \label{Overview}
\end{figure}

Figure \ref{Overview} shows our project design.


\begin{thebibliography}{99}

\bibitem{zhang2023inversion}
Y. Zhang, N. Huang, F. Tang, H. Huang, C. Ma, W. Dong, and C. Xu.
\textit{Inversion-based style transfer with diffusion models}.
In Proceedings of the IEEE/CVF Conference on Computer Vision and Pattern Recognition, pages 10146--10156, 2023.

\bibitem{men2022dct}
Y. Men, Y. Yao, M. Cui, Z. Lian, and X. Xie.
\textit{Dct-net: domain-calibrated translation for portrait stylization}.
ACM Transactions on Graphics (TOG), 41(4):1--9, 2022.

\bibitem{peng2006clip}
Y. Peng and C.-W. Ngo.
\textit{Clip-based similarity measure for query-dependent clip retrieval and video summarization}.
IEEE Transactions on Circuits and Systems for Video Technology, 16(5):612--627, 2006.

\bibitem{liu2021adaattn}
S. Liu, T. Lin, D. He, F. Li, M. Wang, X. Li, Z. Sun, Q. Li, and E. Ding.
\textit{Adaattn: Revisit attention mechanism in arbitrary neural style transfer}.
In Proceedings of the IEEE/CVF International Conference on Computer Vision, pages 6649--6658, 2021.

\bibitem{chen2021pre}
H. Chen, Y. Wang, T. Guo, C. Xu, Y. Deng, Z. Liu, S. Ma, C. Xu, C. Xu, and W. Gao.
\textit{Pre-trained image processing transformer}.
In Proceedings of the IEEE/CVF Conference on Computer Vision and Pattern Recognition, pages 12299--12310, 2021.

\bibitem{karras2020analyzing}
T. Karras, S. Laine, M. Aittala, J. Hellsten, J. Lehtinen, and T. Aila.
\textit{Analyzing and improving the image quality of stylegan}.
In Proceedings of the IEEE/CVF Conference on Computer Vision and Pattern Recognition, pages 8110--8119, 2020.

\bibitem{xiang2017joint}
J. Xiang and G. Zhu.
\textit{Joint face detection and facial expression recognition with mtcnn}.
In 2017 4th International Conference on Information Science and Control Engineering (ICISCE), pages 424--427. IEEE, 2017.

\bibitem{zhang2017beyond}
K. Zhang, W. Zuo, Y. Chen, D. Meng, and L. Zhang.
\textit{Beyond a gaussian denoiser: Residual learning of deep cnn for image denoising}.
IEEE Transactions on Image Processing, 26(7):3142--3155, 2017.

\end{thebibliography}
\end{document}